\documentclass{revtex4}
\usepackage[dvips]{graphicx}

\begin{document}

\title{Heterodyne Near-Field Scattering}
\author{Doriano Brogioli}
\author{Alberto Vailati}
\author{Marzio Giglio}
\affiliation{Dipartimento di Fisica and Istituto Nazionale per la Fisica 
della Materia, Universit\`a di Milano, via Celoria 16,
20133 Milano, Italy}

\begin{abstract}
We describe an optical technique based on the statistical analysis of the 
random intensity distribution due to the interference of the near-field 
scattered light with the strong transmitted beam. It is
shown that, from the study of the two-dimensional power spectrum of the 
intensity, one derives the scattered intensity as a function of the scattering
wave vector. Near-field conditions are specified and discussed.
The substantial advantages over traditional scattering technique are
pointed out, and is indicated that the technique could be of interest
for wave lengths other than visible light.
\end{abstract}

\maketitle

Scattering techniques represent a powerful tool to probe the structure of
matter. Syncrotron radiation, neutron and light scattering allow to
investigate phenomena occurring across a wide range of lengthscales,
spanning from the atomic one up to fractions of a millimeter. 
Scattering is used for the determination of the
structure factor of liquids,
macromolecules, gels, porous materials and complex fluids, 
by detecting the intensity $I_S\left(\Theta\right)$ of the scattered
radiation in the far field as a function of the scattering angle $\Theta$. 
Each point in the far field
is illuminated by radiation coming from different portions of the sample,
and the superposition of the scattered fields 
with random phases gives rise to coherence areas (speckles)
\cite{goodman}. According to Van Cittert and Zernike theorem,
the size and shape of the speckles in the far field is 
related to the intensity distribution of the probe beam.
All the classical techniques rely on the measurement of the average
scattered intensity, and no physical information can be gained from the 
statistical analysis of the far field speckles.

In this paper we will describe a scattering technique based on the statistical
analysis of the random intensity modulation due to the interference of the 
strong transmitted beam and the near field scattered light. We will show that 
one can derive the scattered intensity distribution $I_S\left(\Theta\right)$
from the two dimensional power spectrum of the intensity fluctuations.
The Heterodyne technique is a more powerful and simpler alternative to the 
Homodyne Near Field Scattering that we have recently presented
\cite{carpineti2000,carpineti2001}. It offers many substantial
advantages over the homodyne technique and the conventional small angle light
scattering.
There is no need of the rather awkward block of the transmitted beam as in the 
homodyne case, and this makes the layout very simple and with no necessity of
any alignment. It allows rigorous (static) stray light subtraction without
any blank measurement. Also, being a self referencing technique, it allows
to determine absolute differential scattering cross sections. It also has a 
wide dynamic range, since the signal depends on the amplitude of the 
scattered fields and not on the scattered intensities.
Finally, it gives much improved statistical accuracy.
It is worth pointing out that in principle the technique can be used with
other type of scattering, like syncrotron or FEL radiation, or whenever 
coherence properties are adequate to generate speckles.

The following discussion provides the rationale behind the
heterodyne technique.

As in classical light scattering, we send a collimated laser 
beam with wave vector $\vec{k}_0$ through a sample, and our goal
is to measure $I_S\left(Q\right)$, the intensity of the light scattered at
wave vector $\vec{k}_S$ as a function of the transferred
momentum $\vec{Q}=\vec{k}_S-\vec{k}_0$. 
In the technique presented here, this task is accomplished by
measuring and analyzing the intensity of light
in a plane near the cell, 
in the forward scattering direction (see Fig. 1),
\begin{figure}
\includegraphics{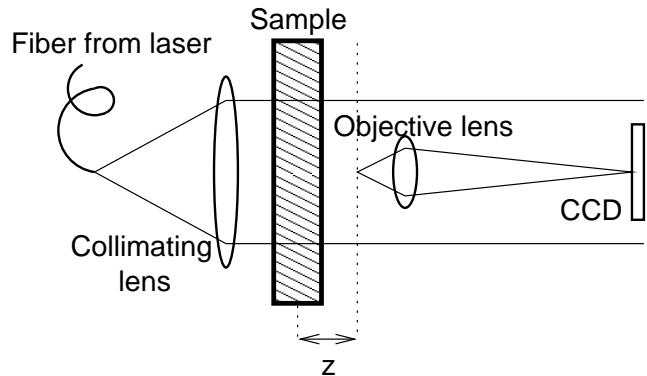}
\caption{
The optical layout of the instrument. The diverging laser beam
from the fiber is collimated and sent through the sample. A microscope
objective images onto a CCD detector a plane at a distance $z$ from the cell.
}
\end{figure}
perpendicular to the direction of
the incident beam, where the intense transmitted light
acts as a reference beam interfering with the weak
scattered beams. Then, in the sensor plane $\left[x,y\right]$,
the intensity $I\left(x,y\right)$ is the sum of the strong
transmitted beam intensity $I_0$ and of the small modulations
$\delta I\left(x,y\right)$, due to the interference of the
transmitted beam with the scattered beams (heterodyne term);
terms arising from the interference between
scattered beams can be neglected (homodyne terms).
The intensity modulation 
$\delta I\left(x,y\right)$, in the plane $\left[x,y\right]$,
can be decomposed in its Fourier components, with amplitude
$\delta I\left(q_x,q_y\right)$. A modulation with wave vector
$\left[q_x,q_y\right]$ is generated by the interference
of the transmitted beam 
with a scattered three-dimensional plane wave with wave vector
$\vec{k}_S=\left[q_x,q_y,k_z\right]$ or 
$\vec{k}_S=\left[-q_x,-q_y,k_z\right]$. Both the waves contribute
to the heterodyne signal:
\begin{equation}
\label{eq_dI_prop_dE}
\delta I\left(q_x,q_y\right) \propto
\Re \left[\delta E\left(q_x,q_y,k_z\right)+
\delta E\left(-q_x,-q_y,k_z\right) \right]
\end{equation}
where $\delta E\left(q_x,q_y,k_z\right)$
and $\delta E\left(-q_x,-q_y,k_z\right)$ are the amplitudes of
the two scattered waves, travelling at symmetric angles with 
respect to the direction of the probe beam.
Because the scattering we are considering
is elastic, the only possible value of $k_z$ can be determined
by imposing the condition 
$\left|\left[q_x,q_y,k_z\right]\right|=k$, where 
$k=\left|\vec{k}\right|$. We can thus easily
evaluate the modulus of the transferred wave vector $Q$,
corresponding to the plane wave responsible for the modulation
of the intensity on the plane, with wave vector $q$:
\begin{equation}
\label{relazione_vettore_trasferito_vettore_misurato}
Q\left(q\right)=
\sqrt{2}k\sqrt{1-\sqrt{1-\left(\frac{q}{k}\right)^2}},
\end{equation}
which can be approximated by $Q\left(q\right)=q$,
in the limit of small scattering angles.
The scattered intensity $I_S\left(Q\right)$
is simply related to the power spectrum
$S_{\delta I}\left(q\right)$ of the intensity modulations in the
detector plane. By taking the mean square modulus of both sides of 
Eq. (\ref{eq_dI_prop_dE}) we obtain:
\begin{equation}
\label{eq_intensita_spettro_potenza_corr}
S_{\delta I}\left(q\right) 
\propto
I_S\left[Q\left(q\right)\right] 
+
\left<
\Re \delta E\left(q_x,q_y,k_z\right) \Re
\delta E\left(-q_x,-q_y,k_z\right)
\right>
.
\end{equation}
Neglecting the correlation term of the two fields, we have:
\begin{equation}
\label{eq_intensita_spettro_potenza}
S_{\delta I}\left(q\right)
\propto
I_S\left[Q\left(q\right)\right].
\end{equation}
In practice the measurement of the intensity modulations is
implemented  by using an array detector, which 
maps the intensity as a function of position $\delta I\left(x,y\right)$.
The evaluation of $S_{\delta I}\left(q\right)$ can then be easily
performed by a suitable processing.

The distance $z$ between the sample and the detector plane
must meet two conditions, that we will discuss below.

Let us consider a sample whose diffraction halo is contained within an angle
$\Theta^*$. The detector is then hit by
light coming from a circular region of the sample with
diameter $D^* \approx z \Theta^*$. 
Equation (\ref{eq_intensita_spettro_potenza}) holds for the ideal case of an 
infinite incoming wave with infinitely wide sample. In order for
Eq. (\ref{eq_intensita_spettro_potenza}) to be valid for a finite size
geometry, we must have that $D^*$ is much smaller than the main beam diameter
$D$ that we assume of uniform intensity. The condition then is 
$z<D/\Theta^*$.

We now discuss the second condition on $z$.
Eq. (\ref{eq_intensita_spettro_potenza}) holds if
$\delta E\left(q_x,q_y,k_z\right)$ and
$\delta E\left(-q_x,-q_y,k_z\right)$ are not correlated.
A typical situation in which this \emph{does not happen} is 
the case of the Raman-Nath scattering regime,
where the sample can be approximated as a two dimensional phase grating,
and the fields scattered at symmetric angles do bear a definite phase
relation. Consequently the system shows the more complex
behaviour described as the Talbot effect \cite{goodman}.
In particular, in spite of the fact
that the grating scatters light at two symmetric angles, at periodic values
of the sensor distance $z$, the interference with the transmitted beam 
does not give rise to any intensity modulation. This is the regime in which
works the shadowgraph technique \cite{cannell1995}.

In order for the two fields not to be correlated,
we must place the sensor not too close to the sample. Since the detector
area $A$ is finite, the
wave vectors $\left[q_x,q_y\right]$ are discrete and correspond to the Fourier 
modes of $A$. Therefore, each mode corresponds to a discrete 
scattering range of directions we can resolve. The light
hitting the area $A$
comes from different regions of the sample, one for each 
scattering direction. In order for 
the phase relations to vanish, 
the distance $z$ must be so large that the portion of the scattering
volume feeding light scattered at 
$\left[q_x,q_y,k_z\right]$ to the detector does not overlap with the portion
of the scattering volume feeding light at
$\left[-q_x,-q_y,k_z\right]$. The resulting condition is $z>L^2/\lambda$,
where $L$ is the transverse dimension of the sensor, and $\lambda$ is the 
wave length of the light used.

For ordinary samples, the mean distance $\delta$ between the scatterers
is such that $\delta \ll L$. In this case, it
can be easily shown that the condition above guarantees that the scattered 
field is a gaussian random process.

To evaluate the performances of the technique we have
performed measurements on water
suspensions of colloidal latex particles;
the wave vector range covered roughly
two decades of $q$ wave vectors.

The optical layout of the instrument is shown in Fig. 1. The collimated and
spatially filtered beam coming from a 10 mW He-Ne LASER impinges onto the
sample contained in a parallel walls cuvette. The beam diameter corresponds
to 21 mm at $1/e$. A 20X microscope objective images onto a CCD detector a
plane placed 15 mm after the cell. The CCD is an array of $768\times 576$
square pixels each having a size of $8\mu \mathrm{m}$. The
intensity distribution onto the CCD is a 20X magnified replica of the
intensity of this pattern. The magnification has been selected so that
the wave vector range we want to measure corresponds to lengths between
the pixel dimension and the sensor dimension.
These magnified speckle patterns
$I\left(x,y\right)$ represent the
raw data, from which the scattered intensity distribution 
$I_S\left(Q\right)$ can be derived according to the the following procedure.
First, a sequence of about 100 images is
grabbed and stored. The images are spaced in time so that the speckle fields
in the images are statistically independent. This is achieved by grabbing
images at a frequency smaller than the smallest characteristic frequency of
the sample, in our case $DQ_{\min }^{2}$, where $D$ is the diffusion
coefficient of the latex particles and $Q_{\min }$ is the smaller wave
vector detected by the CCD ($Q\simeq $20cm$^{-1}$ for our setup). The
time average $\tilde{I}\left(x,y\right) $ of the set of images is then
subtracted from each image, and the result is normalized by the spatial
average $I_{0}$ of $\tilde{I}\left(x,y\right) $: 
\begin{equation}
i\left(x,y\right) =\frac{I\left(x,y\right) -\tilde{I}\left(
x,y\right) }{I_{0}}
\label{segnale_norm}
\end{equation}

Basically $\tilde{I}\left(x,y\right)$ represents an optical background
due to the non-uniform illumination of the sample. This background is
subtracted so to obtain the spatially fluctuating part of the signal. The
spectrum of the normalized signal is then calculated from the Fast Fourier
Transform of the normalized intensity by using Parseval's relation, 
$S(q_x,q_y)=|F[i(x,y)]|^2$. After the azimuthal average of the 
power spectrum,
the scattered intensity is finally obtained from Eq.
(\ref{eq_intensita_spettro_potenza}), where the wave vectors are rescaled
according to Eq. (\ref{relazione_vettore_trasferito_vettore_misurato}).

Data obtained for the intensity scattered in the Mie regime by water
suspension of latex colloidal particles are presented in Fig. 2.
\begin{figure}
\includegraphics[angle=-90]{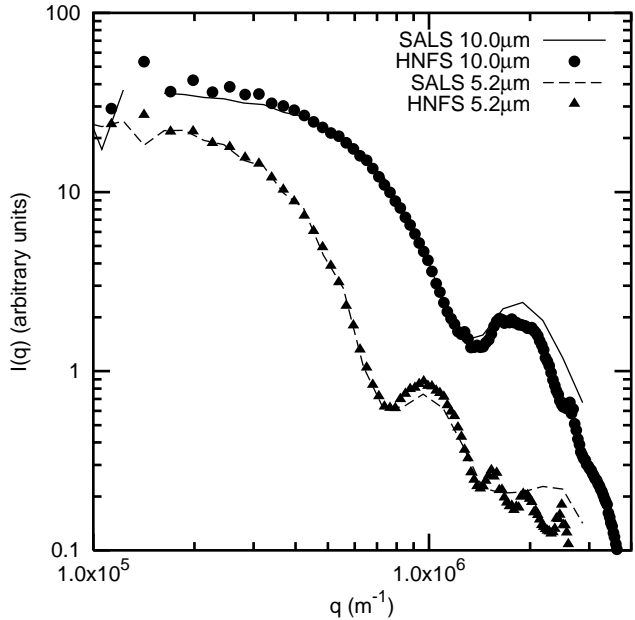}
\caption{
Light scattering from a water suspension 
of $5.2\mu\mathrm{m}$ and $10\mu\mathrm{m}$ latex colloidal particles,
measured by Small Angle Light Scattering (SALS) and Heterodyne Near Field
Scattering.
}
\end{figure}
The two
data-set correspond to 5.2$\mu $m and 10$\mu $m diameter particles.
The concentration was such that the fraction
of power of the probe beam removed due to scattering was of the order of a
few percent, so that the self-beating contribution of the scattered light
is negligeable. Figure 2 also shows data obtained from the same samples by
using a state-of-the-art small-angle light scattering machine
\cite{carpineti1990,ferri1997}
across two decades in wave vector. Data from the heterodyne technique
closely mirror those obtained by means of small angle light scattering.

We believe that visible light Heterodyne Near Field Scattering is a promising
technique, particularly well suited to replace the more traditional low
angle light scattering. Typical applications will include colloids, aggregates,
particulate matter, aerosols, phase transitions and complex fluids in
general. Finally, we point out that X Ray sources like
the FEL should have fairly good coherence properties \cite{TESLA}.
The required magnification to bring the X Ray speckle size to
realistic dimensions and larger than the available pixel sizes 
could be obtained by the use of properly diverging beams
(not discussed here) and long distances between the sample and the sensor.

We thank Marco Potenza for useful discussion.

\end{document}